 \documentclass[12pt,onecolumn]{IEEEtran}

\usepackage{cite}
\usepackage{amsmath,amssymb,amsfonts}
\usepackage{algorithmic}
\usepackage{graphicx}
\usepackage{textcomp}
\usepackage{xcolor}
\def\BibTeX{{\rm B\kern-.05em{\sc i\kern-.025em b}\kern-.08em
    T\kern-.1667em\lower.7ex\hbox{E}\kern-.125emX}}

\usepackage{mathrsfs}
\usepackage{mathtools}
\usepackage{float}
\usepackage{color}
\usepackage{multirow,multicol}
\usepackage{verbatim}
\usepackage{calrsfs}
\usepackage{cancel}
\DeclareMathAlphabet{\pazocal}{OMS}{zplm}{m}{n}

\usepackage[T1]{fontenc}

\usepackage{bm}
\usepackage{cuted,tcolorbox,lipsum}
\definecolor{infocolor}{RGB}{216,237,237}
\usepackage{multicol}
\usepackage{capt-of}
\usepackage{setspace}

\newcommand{\e}[2]{{\mathbb E_#1}\left[ #2 \right]}
\newcommand{\vecc}[1]{{\mathrm {vec}}\left( #1 \right)}
\newcommand{\abs}[1]{\left| #1 \right|}
\newcommand{\T}{T}

\newcommand{\wk}{\bm{\mathrm{w}}_k}
\newcommand{\od}{\omega_{D_d}}
\newcommand{\odi}{\omega_{D_i}}

\newcommand{\Ep}{E_p}
\newcommand{\p}{\bm{\mathrm{p}}}

\newcommand{\ed}{e^{j\od k\T}}

\newcommand{\ei}{e^{j\odi  k\T}}
\newcommand{\eid}{e^{j\dom  k\T}}
\newcommand{\ad}{\alpha_d}

\newcommand{\ai}{\alpha_i }
\newcommand{\sw}{\sigma_w^2 }
\newcommand{\dom}{\Delta\omega }

\newcommand{\zdkn}{\bm{\mathrm{z}}_{d,k}}

\newcommand{\zikn}{\bm{\mathrm{z}}_{i,k}}
\newcommand{\xxi}{\bm{\xi}}

\newcommand{\sumk}{\sum_{k=-\frac{K}{2}}^{\frac{K}{2}}}
\newcommand{\xv}{\bm{\mathrm{x}}}

\newcommand{\iden}{\bm{\mathrm{I}}}
\newcommand{\Ad}{\bm{\mathrm{A}}_d}
\newcommand{\dAd}{\dot{\bm{\mathrm{A}}}_d}
\newcommand{\dAdH}{\dAd^H}
\newcommand{\ddAd}{\ddot{\bm{\mathrm{A}}}_d}
\newcommand{\ddAdH}{\ddAd^H}
\newcommand{\AdH}{\bm{\mathrm{A}}_d^H}
\newcommand{\Add}{\bm{\mathrm{A}}(\theta)}
\newcommand{\Ai}{\bm{\mathrm{A}}_i}

\newcommand{\Aii}{\bar{\bm{\mathrm{A}}}(\theta,\psi)}

\newcommand{\R}{\bm{\mathrm{R}}}

\newcommand{\J}{\bm{\mathrm{J}}}
\newcommand{\Cd}{\bm{\mathrm{C}}_D}

\newcommand{\Rp}{\bm{\mathrm{R}}_p}
\newcommand{\M}{M}
\newcommand{\B}{B}
\newcommand{\art}{\bm{\mathrm{a}}_r}

\newcommand{\artH}{\bm{\mathrm{a}}^H_r}

\newcommand{\dart}{\dot{\bm{\mathrm{a}}}_r}
\newcommand{\att}{\bm{\mathrm{a}}_t}

\newcommand{\datt}{\dot{\bm{\mathrm{a}}}_t}
\newcommand{\Mt}{M_t}
\newcommand{\Mr}{M_r}
\newcommand{\K}{K}

\newcommand{\tr}{\mathrm{tr}}

\newcommand{\adR}{\alpha_{R}}
\newcommand{\adI}{\alpha_{I}}

\begin{document}
\title{{MCRB on DOA Estimation for Automotive MIMO Radar in the Presence of Multipath }}
\author{M. Levy-Israel, {\it Student Member, IEEE}\thanks{Moshe levy-Israel, Igal Bilik, and Joseph Tabrikian are with the School of Electrical and Computer Engineering, Ben Gurion University of the Negev, Beer Sheva, Israel. (e-mails: levyismo@post.bgu.ac.il, bilik@bgu.ac.il, joseph@bgu.ac.il). This work was partially supported by the Israel Science Foundation under Grants 2666/19 and 1895/21.},
I. Bilik, {\it Senior Member, IEEE}, and J. Tabrikian, {\it Fellow, IEEE}
}

\maketitle
\begin{abstract} 
Autonomous driving and advanced active safety features require accurate high-resolution sensing capabilities. Automotive radars are the key component of the vehicle sensing suit. However, when these radars operate in proximity to flat surfaces, such as roads and guardrails, they experience a multipath phenomenon that can degrade the accuracy of the direction-of-arrival (DOA) estimation. Presence of multipath leads to misspecification in the radar data model, resulting in estimation performance degradation, which cannot be reliably predicted by conventional performance bounds. In this paper, the misspecified Cramér-Rao bound (MCRB), which accounts for model misspecification, is derived for the problem of DOA estimation in the presence of multipath which is ignored by the estimator.  Analytical relations between the MCRB and the Cramér-Rao bound are established, and the DOA estimation performance degradation due to multipath is investigated. The results show that the MCRB reliably predicts the asymptotic performance of the misspecified maximum-likelihood estimator and therefore, can serve as an efficient tool for automotive radar performance evaluation and system design. 
\end{abstract}


\section{Introduction}
Recent autonomous driving~\cite{8943254} and active safety technological revolution,~\cite{8633345,7317855}, largely depends on the reliability of the sensing suit and its ability to provide accurate information on the automotive environment~\cite{bilik2019rise,sun2020mimo,8828004,9127857}. 
The sensing suit of the majority of consumer vehicles consists of radars~\cite{Bilik1, Bilik7} and cameras~\cite{9044295,8684792}. While cameras resemble human vision~\cite{7453174,8666747,9307324,8220373}, radars complement them by providing sensing capabilities in poor lighting and harsh weather conditions~\cite{Murtaza,9127853,8828037,8830483}.

Fully autonomous vehicles need to operate in any practical conditions and scenarios, including intersections, urban canyons, tunnels, bridges, and others~\cite{bilik2019rise}. The majority of these scenes are characterized by multipath propagation conditions in both azimuth and elevation~\cite{5606764, 8835603}. Therefore, automotive radars are required to operate also in such multipath-dominated conditions~\cite{6953194,1605750,6494721,6978865} reliably. 

The multipath phenomenon occurs when additional objects are present within the radar field-of-view along with the target~\cite{skolnik2001introduction}. Generally, the signal can be reflected several times from various objects in the transmit path from the radar to the target and in the receive path from the target to the radar. However, in typical automotive scenarios, the indirect paths with multiple reflections are not dominant and can be neglected~\cite{9455253}. This work considers two types of indirect paths. In the first, the multipath only occurs in the transmit path, radar-reflector-target-radar. In the second, the multipath only occurs in the receive path, radar-target-reflector-radar.


The multipath phenomenon may occur in both azimuth and elevation~\cite{6978865}. Automotive radars are typically mounted behind the vehicle bumper and are located close to the road surface. Therefore, automotive radars always operate in the presence of elevation multipath~\cite{6494721},~\cite{9455253,8249143,8249152}. Elevation multipath may also occur in tunnels, below bridges, or over-path road signs and constructions~\cite{9564776}. Horizontal multipath occurs when driving near guardrails, buildings, and adjacent vehicles~\cite{9545546,9636338}. 

The interaction between the electromagnetic wave and the objects experiences specular and diffusive reflections~\cite{9455253}. The wavelength of modern automotive radars operating at $77$ GHz is typically larger than the roughness of the road and the guardrail reflection surfaces. Therefore, the multipath propagation in typical automotive scenarios is mainly induced by specular reflections~\cite{bilik2019rise}. 
Specular reflection is governed by the reflection coefficient, determined by the reflection surfaces, roughness, radar operation frequency, and polarization. All these parameters determine the ratio between the direct and indirect paths in the propagation channel~\cite{skolnik2001introduction}. 

Two possible multipath cases that induce different degradation of the radar target detection and localization performance may occur. 
First, when the direct target echo and the multipath signal are received at different range-Doppler resolution cells, they are processed separately and may result in two separate detections, increasing the probability of false-alarm ~\cite{8835603,8249143}. 
Second, when the direct target echo and the multipath signals appear at the same range-Doppler cell, they are processed jointly. 
In this case, the interference between the direct and multipath signals can be constructive or destructive depending on the phase difference between the direct and indirect paths. In cases when this phase difference induces constructive interference, the multipath increases the signal-to-noise ratio (SNR). Therefore, it improves the target detection range, the probability of detection (PD), and the DOA estimation accuracy. In cases where the phase difference induces destructive interference, the multipath decreases the SNR, degrading the detection range, the PD, and the DOA estimation accuracy. 

Lower bounds on the estimation mean-squared error (MSE) are commonly used to predict radar performance and as a radar system design guideline. In addition to providing a benchmark for performance analysis of estimators, they are frequently used for system design and feasibility study (see e.g. \cite{Athley_TAES,Athley_SSP,Moses_array_design,Li_Tabrikian_Nehorai,Keskin_Koivunen}) 
and in the recent years, they are employed as an optimization criterion in sequential cognitive systems~\cite{Wasim_TSP,Chavali_Nehorai,Haykin_Xue_Setoodeh_Cognitive,Greco_Gini_Bell,Bell_cognitive_2015,Rubinstein_Tabrikian}.
Cram\'{e}r-Rao bound (CRB) is the most widely used bound on the estimation MSE. Its popularity is associated with low computational complexity and asymptotic attainability. However, the CRB ignores model misspecification, such as the presence of multipath. As a result, CRB can not be efficiently used to predict the automotive radar performance degradation in multipath-dominated scenarios. 

Recently,  the misspecified CRB (MCRB) was introduced for radar performance evaluation in scenarios where the model assumed by the estimator is different from the actual model~\cite{richmond2015parameter, vuong1986cramer, fortunati2017performance, ding2011maximum}. The MCRB was extended in several aspects, such as constrained problems \cite{Stefano2}, and applied to various signal processing applications (see e.g.  \cite{Stefano,Zoubir_misspecification}). {The MCRB for range and Doppler estimation in ranging applications, such as globlal navigation satellite systems (GNSS) or single-input single-output (SISO) radar/sonar was introduced in~\cite{lubeigt2023untangling}. }

{In this work, we derive the MCRB for multiple-input multiple-output (MIMO) radar DOA estimation  in the presence of  multipath, which induces model misspecification, that is, the estimator ignores the multipath.} The derived MCRB is used to evaluate the influence of various parameters on the radar performance degradation in the presence of the multipath.
Considering practical scenarios when the automotive radar is mounted close to the road surfaces and experiences the elevation multipath, the effect of indirect path angle, the beam width, and the reflection coefficient (and the resulting ratio between the direct and the indirect paths strengths) on the radar performance is evaluated. 
This work extends our preliminary results in~\cite{OurSAM} in multiple aspects. First, it presents the complete MCRB derivation. Next, it analyzes MCRB properties and evaluates its applicability to practical automotive scenarios. It establishes the relation between the derived MCRB and the conventional CRB. Finally, it provides insight into the influence of the sensor array aperture on the DOA estimation performance in multipath-dominated scenarios.

The main contributions of this work are the derivation of the MCRB for multipath-dominated scenarios and its use for evaluating the radar performance degradation induced by the multipath in typical automotive scenarios. This work demonstrates how the proposed MCRB-based approach for radar performance evaluation unveils dependencies of the radar parameters on the considered scenario and multipath conditions that can not be obtained using the conventional CRB. 
The significance of the derived MCRB is twofold. First, it can be used for automotive radar performance evaluation in typical automotive scenarios in the presence of multipath. Second, the MCRB-based tool can be efficiently used to predict realistic radar performance and assess the performance sensitivity to various radar parameters. Therefore, it can be used for new radars design and their parameters optimization. 

The following notations will be used throughout the paper. Roman boldface lower- and upper-case letters denote vectors and matrices, respectively.  Non-bold italic letters denote scalars. The identity matrix of size $N \times N$ is denoted by $ \mathbf{I}_N $ and $\e{f}{\cdot}$, denotes the expectation w.r.t. the probability density function (PDF), $f(\cdot)$. $\vecc{\cdot}$, $(\cdot)^T$, $(\cdot)^H$, and $\tr(\cdot)$ stand for vectorization, transpose, Hermitian transpose, and trace operators, respectively. 
$\mathrm{Re}\{\cdot\}$ and $\mathrm{Im}\{\cdot\}$ are the real and imaginary operators, respectively.
$\dot{(\cdot)}$ and $\ddot{(\cdot)}$ denote the first and second-order derivative operators, respectively. 
The gradient of the $N \times 1$ vector $\bm{\mathrm{a}}(\bm{\theta})$ w.r.t. the $M \times 1$ vector $\bm{\theta}$ is given by the $N \times M$ matrix, whose $n,m$-th element is: $\left[\nabla\bm{\mathrm{a}}(\bm{\theta})\right] _{n,m}
\triangleq \frac{\partial a_n(\bm{\theta})}{\partial {\theta}_m}$. 

The remainder of this paper is organized as follows. Section II describes the multipath problem formulation. Section III derives the MCRB for the considered problem. Section IV, evaluates the MCRB considering various scenes' geometrical and physical properties and demonstrates its applicability for the radar performance evaluation in typical automotive scenarios. Section V provides a discussion of the main results in the paper and the proposed MCRB-based tool. Section VI summarizes our conclusions.

\section{Problem Formulation}
Consider a mono-static, MIMO radar \cite{Bekkerman_Tabrikian_MIMO} with two colocated arrays of $\Mr$ receivers and $\Mt$ transmitters. The vector of signals transmitted by the transmit array elements is a sequence of $\K$ identical frequency-modulated constant-power baseband signals, $\p(t)$, transmitted at a constant pulse repetition interval (PRI), $\T$. 
In the presence of multipath, the received radar echo consists of two components: a) the direct path radar echo and b) the indirect path echo reflected from various objects in the scene.
Let $\zdkn$ denote the direct path signal component  of the $k$-th pulse/chirp, reflected from a {single-point} target at range, $r_d$, with Doppler angular frequency, $\od$, and DOA, $\theta$:
\begin{equation}\label{echoModel1}
        \zdkn(t)\triangleq{\ad\ed\Add\p(t-\tau_d)},\;\;\; t\in[0, T],
\end{equation}
where $\ad$ is the complex coefficient of the direct path, which includes the two-way propagation loss and phase, $\Add=\bm{\mathrm{a}}_r(\theta)\bm{\mathrm{a}}_t^T(\theta)$ is the MIMO steering matrix to the angle $\theta $, and $\tau_d = \frac{2r_d}{c}$, in which $c$ denotes the propagation speed. In \eqref{echoModel1}, it is assumed that the phase due to Doppler frequency shift is constant within the pulse, and its change is discrete between the pulses. 
{The model in (\ref{echoModel1}) refers to a point target, which can be extended to distributed targets by superposition of the contributions of all the target segments. However, thanks to the high range and Doppler resolution in typical automotive radars, the angular spread in each range-Doppler cell is limited and thus, we can adopt the point target model at each range-Doppler cell.}

\begin{figure}[htp]
    \centering
    \includegraphics[width=8cm,height=8cm,keepaspectratio,]{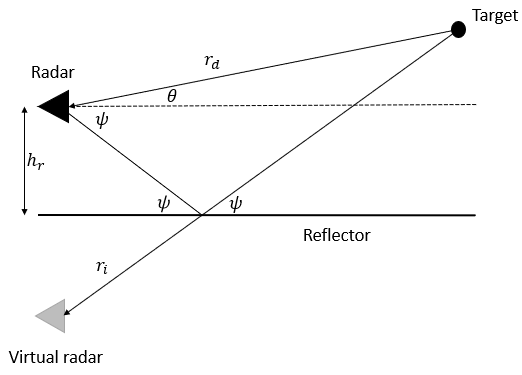}
    \caption{Representative automotive multipath scenario with a target located at the range, $r_d$, with DOA, $\theta$, and a single dominant reflector, at an angle, $\psi$. }
    \label{fig:geo}
\end{figure}

{We will consider the following single-reflected paths:} radar-reflector-target-radar and radar-target-reflector-radar, while neglecting paths with multiple reflections. Considering scenarios with a single dominant reflector (e.g. the ground reflection in Fig. \ref{fig:geo}), the indirect received radar echo consists of two components. The first results from the radar transmission towards the target at DOA $\theta$, and echo received from the reflector at DOA $\psi$ with the corresponding MIMO steering matrix, $\bm{\mathrm{a}}_r(\psi)\bm{\mathrm{a}}_t^T(\theta)$. The second results from the radar transmission towards the reflector at DOA $\psi$, and the echo received from the target at DOA $\theta$, with the corresponding MIMO steering matrix, $\bm{\mathrm{a}}_r(\theta)\bm{\mathrm{a}}_t^T(\psi)$. Since both paths have the same physical properties, such as length, reflections material, and angles, they share the same propagation loss coefficient. Therefore, the indirect path echo can be modeled as follows:
\begin{equation}
        \zikn(t)\triangleq{\ai\ei\Aii\p(t-\tau_i)}\;,
\end{equation} 
where $\tau_i=\frac{2r_i}{c}$\ is the indirect path delay, $\ai $ is the indirect path complex coefficient, which includes the indirect propagation loss and phase, $\odi$ is the indirect path Doppler angular frequency, and $\Aii \triangleq{\bm{\mathrm{a}}_r(\psi)\bm{\mathrm{a}}_t^T(\theta)+\bm{\mathrm{a}}_r(\theta)\bm{\mathrm{a}}_t^T(\psi)}\;.$

The received signal $\bm{\mathrm{x}}_k$ can be modeled as a superposition of the direct and indirect paths along with additive noise: 
\begin{equation}\label{measModel1}
\bm{\mathrm{x}}_k(t)=\zdkn(t) + \zikn(t) +\wk(t)\;, 
\end{equation} 
where $\{\wk(t)\}$ is an additive Gaussian noise, assumed to be white in fast time, $t$, and slow time, $k$, with known covariance  matrix, $\mathbf{R}$.  The phases of the complex path loss coefficients depend on the reflector range and the reflection coefficient phase: \begin{align}\label{eqad}
    &\ad = \alpha_{0,d}|\Gamma_t|e^{j(\angle \Gamma_t+\phi_{r_d})}\;,\\
    &\ai = \alpha_{0,i}|\Gamma_t|\cdot|\Gamma_r|e^{j(\angle\Gamma_t+\angle\Gamma_r+\phi_{r_i})}\label{eqai}\;,
\end{align}
where {$\alpha_{0,d}$ and $\alpha_{0,i}$ are the propagation loss coefficients in the direct and indirect paths, respectively, $\Gamma_t$ and $\Gamma_r$ are the reflection coefficient of the target and the reflector surface, respectively}. The range residual phase components are $\phi_{r_d}=\frac{2\pi r_d}{\lambda}$ and $\phi_{r_i}=\frac{2\pi r_i}{\lambda}$, where $\lambda$ denotes the wavelength. The phase difference between the paths is given by
\begin{equation}
    \Delta \phi = \angle \ad - \angle \ai\;.
    \label{dphi}
\end{equation}
In a multipath-dominated scenario, the ratio between the path loss coefficients $\ad, \ai$ plays a major role in determining the system performance. 
The amplitude ratio is measured via the signal-to-multipath ratio (SMR): 
\begin{equation}\label{SMR}
\mathrm{SMR} = \frac{\abs{\ad}^2}{\abs{\ai}^2}\;.
\end{equation}

For simplicity of the following discussion, the arguments $\psi, \theta$ will be omitted using the following notations:
\begin{align*}
  &\Ai\triangleq{\Aii}\;,\\  
  &\Ad \triangleq{\Add}\;.
\end{align*}

The assumed model ignores the presence of multipath, and the data model does not contain the indirect path echoes $\zikn(t)$. Hence, it can be expressed as
\begin{equation}\label{measModel2}
\bm{\mathrm{x}}_{k}(t)=\zdkn(t) +\wk(t)\;. 
\end{equation}
The noise statistics are assumed to be identical to the case of the true model. Our goal is to evaluate the radar DOA estimation performance using measurements, $\bm{\mathrm{x}}_{k}(t)$, according to the model in \eqref{measModel1}, while the assumed data model is \eqref{measModel2}, and considering the following unknown parameters vector 
\begin{equation}\label{xxi_def}
\xxi=\left[
\mathrm{Re}\left\{\ad \right\} \;\;\mathrm{Im}\left\{\ad \right\}\;\; \tau_d \;\; \od  \;\; \theta \right]^T\;.
\end{equation}
{The proposed MCRB is derived for the case where the estimator ignores the presence of multipath and thus, the vector of unknown parameters does not include the multipath parameters, $\psi, \tau_i, {\omega_D}_i, \alpha_i$. However, the MCRB depends on the true model parameters, as shown in the next section.}

In the next section, we will derive the MCRB for DOA estimation in order to evaluate the effect of model misspecification
due to the presence of a multipath.

\section{MCRB Derivation}

Conventionally, the maximum likelihood (ML) estimator of $\xxi$ is derived from the assumed PDF, $f_{\xv}(\xv;\xxi)$, which may be different from the actual PDF $g_{\xv}(\xv)$. For this problem, the MCRB provides a lower bound on the MSE of estimators, with bias obeys the asymptotic bias of the misspecified ML (MML) estimator, and it is given by ~\cite{richmond2015parameter,vuong1986cramer,Fortunati,ding2011maximum}: 
\begin{align}\label{MCRB1}
    &\e{g}{(\hat{\xxi}-\bm{\mathrm{\xxi}})(\hat{\xxi}-\bm{\mathrm{\xxi}})^T}
    \succcurlyeq\\\nonumber&
    {{\Cd^{-1}(\bm{\xxi})\J_{(g,f)}(\bm{\xxi})\Cd^{-1}(\bm{\xxi})}}+{(\xxi-\bm{\xi}_A)(\xxi-\bm{\xi}_A)^T}\;,
  \end{align}
where the first term in the right-hand side of (\ref{MCRB1}) represents the covariance contribution, while the second term represents the model misspecification-induced bias. In~\eqref{MCRB1}, $\mathbf{J}_{(g,f)}(\xxi)$ is the misspecified Fisher information matrix (FIM):
\begin{equation}
    \mathbf{J}_{(g,f)}(\xxi) = \e{g}{\left(\frac{\partial \mathrm{log} f_{\xv}(\xv;\xxi)}{\partial\xxi}\right)^T\frac{\partial \mathrm{log} f_{\xv}(\xv;\xxi)}{\partial\xxi}}\;,
\end{equation}
and $\bigg[\Cd(\xxi)\bigg]_{ij}$ is the $(i,j)$th element of the error-score function:
\begin{equation}
        \bigg[\Cd(\xxi)\bigg]_{ij}\triangleq\\ \e{g}{\left(\frac{\partial^2 \mathrm{log} f_{\xv}(\xv;\xxi)}{\partial\xi_i\xi_j}\right)}\;,
\end{equation}
where $\xxi_A$ denotes the ML convergence bias term under model misspecification: 
\begin{equation}\label{kld}
    \hat{\xxi}_{MML}\rightarrow \mathrm{arg}\;\underset{\xxi'}{\mathrm{min}}\; \mathrm{KLD}(g_{\xv}(\xv)\| f_{\xv}(\xv,\xxi'))\triangleq{\xxi_A}\;,
\end{equation}
where $\mathrm{KLD}(\cdot \| \cdot)$ denotes the Kullback-Leibler divergence. 

In the multipath-free scenario, the MML estimator is derived assuming the direct path only. The additional indirect paths are considered model misspecification. For Gaussian distribution, i.e. $f_{\xv}(\xv;\xxi)={\pazocal{CN}(\bm{\mathrm{\mu}}(\xxi), \R)}$, and, $g_{\xv}(\xv)={\pazocal{CN}}(\bm{\mathrm{\mu}}',\R')$, {the misspecified FIM (MFIM), $\J_{(g,f)}(\xxi)$}, can be calculated by the generalized Slepian-Bangs formula ~\cite{richmond2015parameter}:
\begin{equation}\label{gslep}
    \J_{(g,f)}(\xxi)\triangleq{2\mathrm{Re}\left\{{\nabla\bm{\mathrm{\mu}}^T(\xxi)\R^{-1}\R'\R^{-1}\nabla\bm{\mathrm{\mu}}(\xxi)}\right\}}\;,
\end{equation}
and the matrix $\mathbf{C}_D$  can be calculated as:
\begin{equation}\label{CDdef}
   \bigg[\Cd(\xxi)\bigg]_{ij}\triangleq\\ \left[\J(\xxi)\right]_{ij}- 2\mathrm{Re}\left\{\frac{\partial^2\bm{\mathrm{\mu}}^T(\xxi)}{\partial\xi_i\partial\xi_j}\R^{-1}\Delta\bm{\mathrm{\mu}}\right\}\;,
\end{equation}
where $\Delta\bm{\mathrm{\mu}}$ is
\begin{equation}\label{deltaMu}
    \Delta\bm{\mathrm{\mu}}\triangleq{\bm{\mathrm{\mu}}'-\bm{\mathrm{\mu}}(\xxi)}\;.
\end{equation}
For the derivation of the bound, we assume that under both the true and the assumed models, the data samples over time are statistically independent. The assumed data distribution for a given time instance of the data vector is
$\bm{\mathrm{x}}_k(t) \sim{\pazocal{CN}(\zdkn(t), \mathbf{R})}$ where the true distribution of the data vector is $\bm{\mathrm{x}}_k(t) \sim{\pazocal{CN}(\zdkn(t)+\zikn(t), \mathbf{R})}$.
Since the misspecification is in the mean only, then, $\R' = \R$, {and the MFIM in \eqref{gslep} coincides with the conventional FIM, i.e,} $\J_{(g,f)}(\xxi) = \J(\xxi)$, where $\J(\xxi)$ stands for the conventional FIM. 

The resulting MCRB for estimating $\xxi$ is given by
\begin{align}
    &\e{g}{(\hat{\xxi}-\xxi)(\hat{\xxi}-\xxi)^T}\succcurlyeq\\\nonumber &\Cd^{-1}(\bm{\xxi})\J(\bm{\xxi})\Cd^{-1}(\bm{\xxi})+(\xxi-\xxi_A)(\xxi-\xxi_A)^T=\mathbf{M}+\mathbf{B}\;, \nonumber
\end{align}
where the covariance matrix, $\mathbf{M}$, and the bias contribution to the MSE, $\mathbf{B}$, are
\begin{align}
    &\mathbf{M} \triangleq \Cd^{-1}(\bm{\xxi})\J(\bm{\xxi})\Cd^{-1}(\bm{\xxi})\;,\\
    &\mathbf{B} \triangleq (\xxi-\xxi_A)(\xxi-\xxi_A)^T\;.
    \end{align}
The derivation of the elements of the matrices $\J(\bm{\xxi})$ and $\mathbf{C}_D(\xxi)$ appears in Appendices A and B, respectively. 

In this paper, we are interested in DOA estimation performance. Therefore, an analytic expression for the corresponding $\theta$ element in the diagonal of $\mathbf{M}$, denoted as $M_{\theta \theta}$, is derived in the following. 

The multipath effect on the DOA estimation is expected when both the direct and indirect paths, induced by the target and the reflector, fall within the same range-Doppler cell: 
\begin{align}
    &\Delta r = r_i-r_d < R_{res}\;,\\
    &\dom = \odi-\od < \omega_{D,res}\;,
\end{align}
where $R_{res}$ and $\omega_{D,res}$ are the radar range and Doppler resolution, respectively.  The exact position of $\Delta r$ and $\Delta \omega$ within the range-Doppler cell affects the power ratio and phase difference between the paths. However, since we are interested in DOA estimation analysis in a given range-Doppler cell, this effect can be neglected for the $\M_{\theta\theta}$ derivation by assuming:
\begin{align}
    &\Delta r \ll R_{res}\label{ass1}\;,\\
    &\dom \ll \omega_{D,res}\label{ass2}\;.
\end{align}
In addition, the data can be spatially whitened, therefore, without loss of generality, it can be assumed that 
\begin{equation}
    \mathbf{R}=\mathbf{\sw \iden}_{M_r}\;.\label{AssumR}
\end{equation} 
where $\iden_{M_r}$ is the identity matrix of size $M_r$. 
Appendix C shows that by using assumptions \eqref{ass1}-\eqref{AssumR}, the MCRB can be expressed as:
\begin{equation}
    \mathrm{MCRB}(\theta)=\M_{\theta\theta}+\B_{\theta\theta}\;,\label{MCRB}
\end{equation}
where
\begin{align}\label{cd1}
    &\M_{\theta\theta}=\mathrm{CRB}(\theta)\frac{E_{\Dot{A}}\left(\left|\tr(\dAdH\Ai )\right|^2+\mathrm{SMR} E_{\Dot{A}}\right)}{\left(\mathrm{Re}\left\{ \tr(\ddAdH\Ai)e^{-j\Delta \phi}\right\}-\sqrt{\mathrm{SMR}}E_{\Dot{A}}\right)^2}\\\nonumber\;\\
    &\B_{\theta\theta}=(\theta - \theta_{A})^2\;.\label{BIAS_form}
\end{align}
The KLD in (\ref{kld}) is minimized by $\theta'$ that maximizes the inner product between $\zdkn$ and $\zikn$ as:
\begin{align}\label{cd2}
    \theta_A = & \mathrm{arg}\;\underset{\theta'}{\mathrm{max}}\;
    \bigg\{|\tr\left(\mathbf{A}^H(\theta')\mathbf{A}(\theta)\right) + \nonumber \\ &\frac{\ai}{\ad+\ai}\tr\left(\mathbf{A}^H(\theta') \bar{\mathbf{A}}(\theta,\psi)\right)|^2\bigg\}\;.
\end{align} 
The FIM for estimating $\xxi$, based on the data model in \eqref{measModel2} is derived in Appendix A, and the CRB for estimating $\theta$ is given by 
\begin{equation}\label{CRB}
    \mathrm{CRB}(\theta)=\frac{1}{\mathrm{J}_{\theta\theta}}=\frac{1}{2\mathrm{SNR}}\frac{1 }{K\Ep E_{\Dot{A}}}\;,
\end{equation}
where $E_p$ is the chirp energy, $\J_{\theta\theta}$ is the corresponding FIM component in~\eqref{Jxi2}, $E_{\Dot{A}}$ is defined as:
\begin{equation}
  E_{\Dot{A}}\triangleq{\tr(\dAd\dAdH)} = \|\Dot{\bm{\mathrm{a}}}_r\|^2 + \|\Dot{\bm{\mathrm{a}}}_t\|^2\;,
\end{equation}
and  
\begin{equation}
\label{SNR_def}
    \mathrm{SNR} = \frac{\abs{\ad}^2}{\sw}\;.
\end{equation}
In the following, the MCRB is analyzed in two extreme cases. First,  when the indirect path amplitude equals zero, $\ai=0$, the  $\mathrm{SMR}\to\infty$, and the MCRB is expected to coincide with the CRB:
\begin{align}
      &\lim_{\mathrm{SMR}\to\infty} \frac{M_{\theta \theta}}{\mathrm{CRB}(\theta)}=\\&\lim_{\mathrm{SMR}\to\infty}\frac{E_{\Dot{A}}\left(\left|\tr(\dAdH\Ai )\right|^2+\mathrm{SMR} E_{\Dot{A}}\right)}{\left(\mathrm{Re}\left\{ \tr(\ddAdH\Ai)e^{-j\Delta \phi}\right\}-\sqrt{\mathrm{SMR}}E_{\Dot{A}}\right)^2}=1\;.\nonumber
\end{align}
In addition, \eqref{cd2} results in $\theta_A=\theta$, and from \eqref{BIAS_form}, $B_{\theta\theta}=0$. Therefore, in this case,  \eqref{MCRB}, can be simplified as:
\begin{align*}
     \mathrm{MCRB}(\theta)=\mathrm{CRB}(\theta)\;.
\end{align*}
In the second case, when all the three paths are coherent, $\ai=\ad$, i.e., $\mathrm{SMR} = 1$ and $\psi=\theta$, the amplitude of the received radar echo is tripled since it consists of the direct path and two indirect paths. The bias term $B_{\theta \theta} = 0$, and the CRB in  $(\ref{CRB})$ is reduced by the factor $\frac{1}{9}$ as follows:
\begingroup
\allowdisplaybreaks
\begin{align}
\label{MCRB_equal_paths}
    \mathrm{MCRB}(\theta) &=\M_{\theta\theta}\nonumber \\&=\mathrm{CRB}(\theta)\frac{E_{\Dot{A}}\left(\left|\tr(\dAdH\Ad )\right|^2 +E_{\Dot{A}}\right)}{\left(2\mathrm{Re}\left\{ \tr(\ddAdH\Ad)\right\}-E_{\Dot{A}}\right)^2}  \nonumber  \\
  &   \stackrel{(a)}{=} \mathrm{CRB}(\theta)\frac{E_{\Dot{A}}^2}{\left(2\mathrm{Re}\left\{ \tr(\ddAdH\Ad)\right\}-E_{\Dot{A}}\right)^2} \nonumber  \\
    &\stackrel{(b)}{=}\mathrm{CRB}(\theta)\frac{E_{\Dot{A}}^2}{\left(-3E_{\Dot{A}}\right)^2} = \frac{1}{9}\mathrm{CRB}(\theta)\;,
\end{align}
\endgroup
where (a) follows from the considered symmetry of the transmit and receive arrays with respect to a common center, $\artH\dart=\datt\att = 0$, which leads to:
\begin{equation}\label{i3}
    \tr(\dAd\AdH)=\mathrm{tr}((\dart\att^T+\art\datt^T)\att^*\art^H)=0\;,
\end{equation}
and (b) can be obtained using \eqref{i3} and the following identity, obtained by a few lines of algebra: 
\begin{equation}\label{i4}
    \tr(\ddAdH\Ad)=-\mathrm{tr}(\dAd\dAdH)=-E_{\Dot{A}}\;.
\end{equation}
\section{Numerical MCRB Evaluation}
This section first investigates the properties of the derived MCRB, and next demonstrates its use in radar performance evaluation in practical automotive scenarios.
\subsection{Numerical Analysis}
\begin{figure}[htp]
    \centering
    \includegraphics[width=8cm,height=6cm]{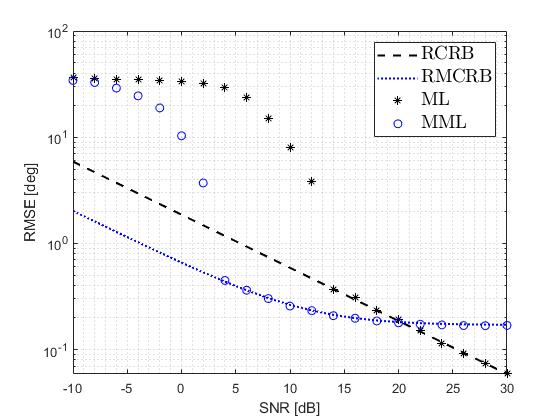}
    \caption{DOA estimation RMSE of ML and MML, compared to the RMCRB and RCRB on the RMSE of the target DOA estimation as a function of SNR. The target DOA is $\theta=0^\circ$ and the indirect path direction is at $\psi=0.5^\circ$ with SMR=$0$ dB and equal phases, $\Delta \phi =0$.}
    \label{fig:ML}
\end{figure}

In this subsection, we consider a MIMO radar of with $M_t = 3$ and $M_r = 4$ transmit and receive elements, respectively. The receive array is a uniform linear array (ULA) with half wavelength inter-element spacing, and the transmit array is a sparse ULA, whose inter-element spacing is equal to the receive array aperture, such that the virtual MIMO array is a ULA of $M_tM_r=12$ elements with half a wavelength inter-element spacing. Considering the multipath scenario, the radar observes a target at DOA, $\theta=0^\circ$, and the indirect path at DOA, $\psi=0.5^\circ$, with SMR of 0 dB and equal phases, $\Delta \phi=0$. Fig. \ref{fig:ML} shows the derived root-MCRB (RMCRB) and the root-CRB (RCRB) compared to the evaluated root-mean-squared-error (RMSE) of the MML and ML as a function of SNR. Both estimators assume the model in \eqref{measModel2}, but for the MML estimator, the data is generated based on \eqref{measModel1}. Fig. \ref{fig:ML} demonstrates the ability of the RMCRB to predict the MML RMSE of the DOA estimation. The RMSE of the MML is limited due to the dominant bias at high SNRs. { The contribution of the bias term in the RMCRB is more significant at higher SNRs and can dominate the variance term. This behavior causes the RMCRB to flatten with increasing SNR.} Notice that at SNRs below $20$dB, the RMCRB is lower than the RCRB. This phenomenon can be explained by the constructive interference between the direct and indirect paths when the difference between the direct and indirect path DOAs is small, $\psi \approx \theta$, which increases the receiver SNR and improves the estimation accuracy. 

\begin{figure}[htp]
    \centering
    \includegraphics[width=8cm,height=6cm]{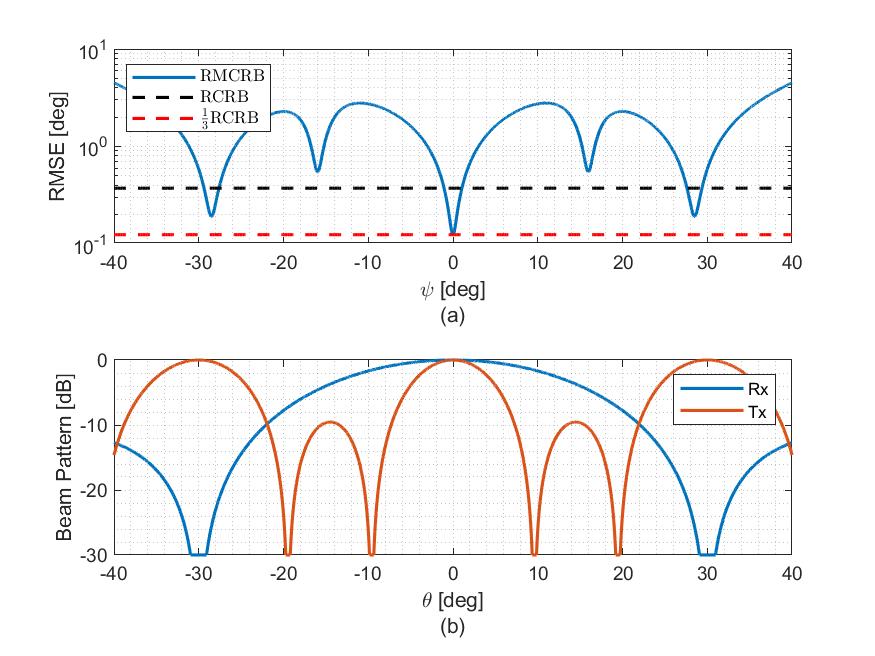}
    \caption{(a) RMCRB and RCRB on the DOA estimation RMSE as a function of the difference between the direct and indirect path DOAs, $\Delta\theta$, for target DOA, $\theta=0^\circ \;  (\Delta\theta=-\psi)$, SMR=$0$ dB and equal phases, $\Delta \phi =0$. (b)  Beampatterns of the transmit and receive arrays in with $M_t = 3$ and $M_r = 4$ elements.}
    \label{fig:MCRB3}
\end{figure}

\begin{figure}[ht]
    \centering
\includegraphics[width=8cm,height=6cm]{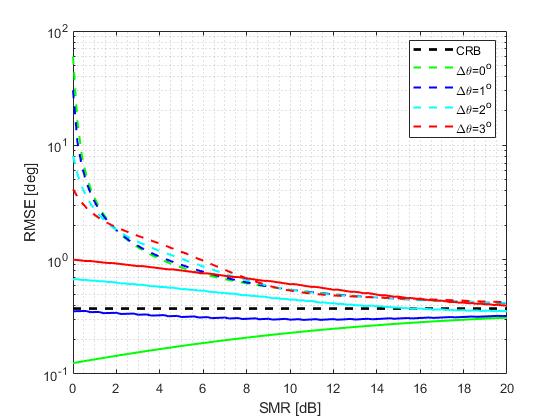}
    \caption{MCRB and CRB on the  DOA estimation RMSE as a function of SMR, with target DOA, $\theta=0^\circ$ and SNR = $10$ dB. {Solid and dashed lines represent constructive  ($\Delta \phi = 0$) and   destructive ($\Delta \phi = |2\pi/3|$) multipath scenarios, respectively.}} 
    \label{fig:MCRB4}
\end{figure}

\begin{figure}[htp]
    \centering
    \includegraphics[width=11cm,height=9cm]{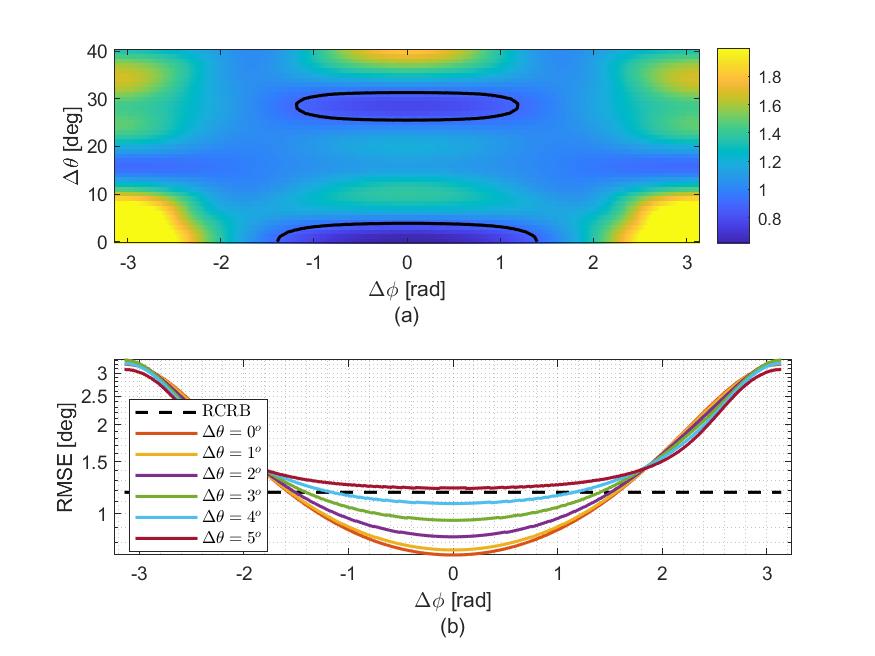}
    \caption{ MCRB and CRB on target DOA estimation as a function of the phase difference between the direct and indirect paths, $\Delta\phi$, with target DOA, $\theta=0^\circ$, SNR = $10$ dB, and SMR = $10$ dB. {(a) The ratio between the RMCRB and the RCRB. (b) Slices of RMCRB as a function $\Delta\phi$}.}
    \label{fig:MCRB6}
\end{figure}

The RMCRB on the target DOA estimation is affected by the DOA difference, $\Delta\theta \triangleq \theta - \psi$, phase difference defined in (\ref{dphi}), and power ratio defined in (\ref{SMR}), between the direct and indirect paths. Figs.~\ref{fig:MCRB3}-\ref{fig:MCRB6} investigate the influence of these terms on the RMCRB, in fixed SNR conditions. 

Fig. \ref{fig:MCRB3} shows the RMCRB sensitivity to the difference between the direct and indirect DOAs, $\Delta\theta$. Subplot (a) shows the RMCRB in the case where the phase difference between the direct and indirect paths is $\Delta\phi=0^\circ$. Subplot (b) shows the transmit and receive beampatterns of the considered array of the MIMO radar. Notice that the multipath effect on the RMCRB on DOA estimation RMSE in subplot (a) is negligible at low $\Delta\theta$ and grows with increasing the angle difference within the main lobe angles. The minimal RMSE value, $\frac{1}{3}{\mathrm{RCRB}}$, is obtained when $\psi = \theta$, as expected since the indirect paths are constructive and the amplitude of the target echo is tripled. Notice the correspondence between the TX array beampattern grating lobes at $\psi = \pm 30^\circ$ in subplot (b) and the local minima of the RMSE in subplot (a). Although there is a null in the RX beampattern at $30^\circ$, the indirect path contribution is obtained through the transmitters, which eventually improves the DOA estimation accuracy. 

Fig.~\ref{fig:MCRB4} shows the multipath influence on the RMCRB on the target DOA estimation as a function of the SMR for various angle differences between the target and the reflector DOAs, $\Delta\theta$. The solid lines represent the scenario where the phases of the direct and indirect paths are equal, $\Delta\phi=0$. The dashed lines represent the scenario where the direct path and the two indirect paths are destructive, i.e., $\Delta\phi=\frac{2\pi}{3}$. (as it is the required phase difference between 3 equal amplitude vectors to be destructive). 
Notice that at high SMR, the direct path is dominant, and therefore, the multipath propagation does not affect the target DOA estimation RMSE, which converges to the RCRB. However, at low SMR, the indirect path becomes dominant, and the RMSE is determined by the bias, $\theta_A$, which increases with increasing $\Delta\theta$. Interestingly, at low SMR and small $\Delta\theta$ values, the multipath is constructive, and the RMCRB on the target DOA estimation RMSE is lower than the RCRB. Notice that at low SMRs, the RMCRB sensitivity to the DOA angle difference, $\Delta\theta$, is higher due to the bias term dominance in this area. 
Finally, notice the {high RMSE values} of the dashed lines around SMR=$0$dB, where complete destruction between paths occurs. The values of these peaks decrease with increasing $\Delta \theta$. 

{Increasing the SNR (for example, by higher transmit power or larger RCS) will equally increase the power of the direct and indirect echoes, such that the SMR which determines the DOA performance will remain constant, and hence for a given destructive interference phase difference, i.e.,  $\Delta\phi = |\frac{2\pi}{3}|$, the DOA performance will not improve. Of course that theoretically, increasing the SNR without affecting the indirect echo power will increase the SMR, and as a result, the DOA estimation performance will improve.}

{Subplot (a) in Fig.~\ref{fig:MCRB6} presents the ratio between RMCRB and RCRB as a function of phase difference, $\Delta\phi$, and  DOA difference, $\Delta\theta$, between the direct and indirect paths. In the areas within the black contours, the RMCRB is lower than the RCRB. Interestingly, when the phase difference between the target and the indirect path DOAs is small, $|\Delta\phi|<1.5$ radians, and $\Delta\theta$ is less than 6$^o$, the multipath improves the DOA estimation performance, and the RMCRB is lower than the RCRB. Similar behavior can be observed when $\Delta\theta\approx 30^o$. This observation can be explained by grating lobes in the beampattern at $\theta=30^\circ$, causing one of the indirect paths to leak into the received signal and potentially improve the DOA estimation. Subplot (b) in Fig.~\ref{fig:MCRB6} shows that as $|\Delta\theta|$ increases, the RMCRB's sensitivity to the phase difference decreases, resulting in flatter curves along $\Delta\phi$. This phenomenon is attributed to the fact that when the indirect path experiences lower array gain in the beampattern, the effect its phase, of $\Delta\phi$, is lower.}

\subsection{MCRB for Automotive Scenario}
\begin{figure}[htp]
    \centering
    \includegraphics[width=0.45\textwidth]{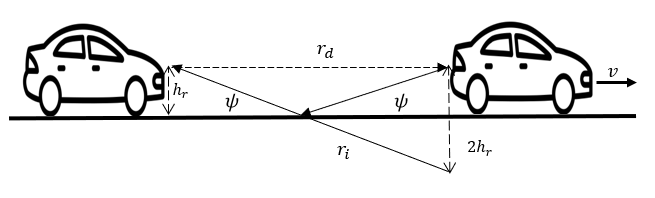}
    \caption{A typical automotive scenario where a front radar, mounted at the height, $h_r$, on the host vehicle, detects a target vehicle at the range, $r_d$, moving with relative velocity, $v$, on the flat horizontal road surface (elevation angle $\theta=0^\circ$). The ground multipath induces two indirect paths of length, $r_i$, at angle $\psi$.}  
    \label{fig:scenario}
\end{figure}

This subsection demonstrates the ability of the derived RMCRB to predict automotive radar performance in practical, multipath-dominated automotive scenarios. Consider a typical automotive scenario in Fig. \ref{fig:scenario}. The front radar on the host vehicle (left vehicle) detects a moving target vehicle at the range, $r_d$. Let the radar range resolution be $R_{res} = 0.5$ m and Doppler resolution $D_{res}=0.05$ m/s.

The indirect path length is given by 
\begin{equation}\label{ri}
    r_i=\sqrt{r^2_d\cos^2\theta+(r_d\sin\theta+2h_r)^2}\;,  
\end{equation}
where $h_r$ is the radar height from the flat road surface.
The indirect path angle $\psi$ is given by
\begin{equation}\label{psi_}
    \psi = \cos^{-1}\left(\frac{r_d\cos\theta}{r_i}\right)\;.
\end{equation}

This scenario considers a multipath reflection from a dry asphalt road surface with the following vertical complex reflection coefficients ~\cite{mahafza2005radar}: 
\begin{equation}\label{gamma_ver}
    \Gamma_r = \frac{\epsilon\sin\psi-\sqrt{\epsilon-\cos^2\psi}}{\epsilon\sin\psi+\sqrt{\epsilon-\cos^2\psi}}\;,
\end{equation}
where 
\begin{equation}
     \epsilon = \epsilon_r -j60\lambda\cdot\gamma\;,
\end{equation}
in which $\epsilon_r$ is the road surface dielectric constant, $\gamma$ is the road surface conductivity, and $\lambda$ is the wavelength. 
In typical automotive scenarios, $\epsilon_r = 4$, $\gamma = 0.005$~\cite{ntaikos2017channel}, and automotive radar wavelength is $\lambda = 3.8$ mm. Note that the change in range affects the reflection DOA, $\psi$, which controls the reflection coefficient,  $\Gamma_r$, and thus, the path loss coefficient, $\ai$. The asymptotic behavior of the scenario parameters in Fig.~\ref{fig:scenario} shows that the difference between the direct and indirect path DOAs, $\Delta \theta$, diminishes with increasing the target range, $r_d$, and when $\theta = 0^\circ$: 
\begin{equation}\label{psiasympt}
  \lim_{r{_d}\to\infty} \psi=-\lim_{r{_d}\to\infty} \Delta\theta =0\;.
\end{equation}

From \eqref{gamma_ver} and \eqref{psiasympt}, it can be seen that for long distances, $\psi \approx 0$, and the vertical component of the reflection coefficient, denoted as $\Gamma_r$, represents a perfect mirror:    
\begin{equation}\label{rdasympt}
  \lim_{r{_d}\to\infty} \Gamma_r = -1\;.
\end{equation}
As a result, the indirect path complex coefficient, $\alpha_i$, can be obtained by substituting \eqref{rdasympt} in \eqref{eqai} as:
\begin{align}
  &\lim_{r{_d}\to\infty} \alpha_i = -|\Gamma_t|e^{j(\angle\Gamma_t+\phi_{r_i})}\;,
\end{align}
and the phase difference between the direct and indirect paths is 
\begin{equation}\label{deltaphiasympt}
    \lim_{r{_d}\to\infty} \Delta \phi = \phi_{r_d}-\phi_{r_i}+\pi\;.
\end{equation}
Notice that when $r_d$ grows, the phase difference between the direct and indirect paths, $\Delta \phi$, is determined by the difference between the path lengths, $\Delta r$. 
{The direct and indirect path coefficients $\ad$ and $\ai$ were computed using \eqref{eqad} and \eqref{eqai} at each target range}, 
where $\Gamma_r$ is derived by substituting the reflector DOA $\psi$ in \eqref{gamma_ver}, where $\psi$ is derived using \eqref{psi_}.

\begin{figure}[htp]
    \centering
    \includegraphics[width=8cm,height=6cm]{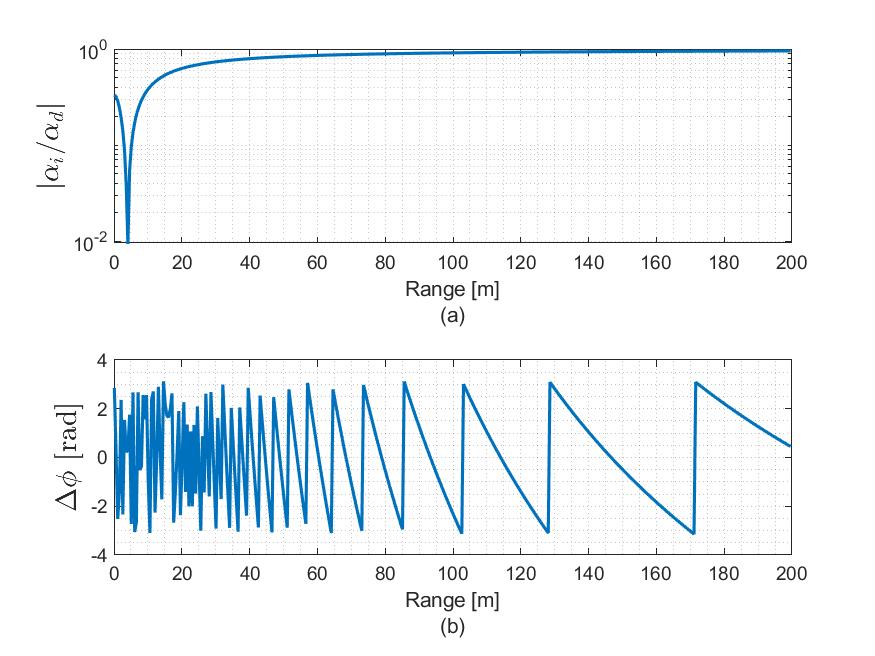}
    \caption{Amplitude ratio and phase difference between the direct and indirect paths, as a function of the radar target range, in a typical automotive scenario in Fig. \ref{fig:scenario}, in the presence of vertical multipath from the road surface.}
    \label{fig:scenario_1}
\end{figure}

Fig.~\ref{fig:scenario_1} shows the normalized amplitude $\left|\frac{\alpha_i}{\alpha_d}\right|$, and the phase difference $\Delta \phi$, in the vertical multipath scenario as a function of range. Subplot (a) in Fig.~\ref{fig:scenario_1} shows the normalized amplitude of the indirect path coefficient. The null obtained at a range of $4$ meters is due to high reflector DOA values at this area, which yields a small amplitude of the reflection coefficient $\Gamma_r$. Notice that the difference between the direct and indirect path lengths, $\Delta r$, varies rapidly at short and slowly at long ranges. As a result, the phase difference, $\Delta \phi$ in subplot (b), fits the result in~\eqref{deltaphiasympt}. 

\begin{figure}[htp]
    \centering
    \includegraphics[width=9cm,height=9cm]{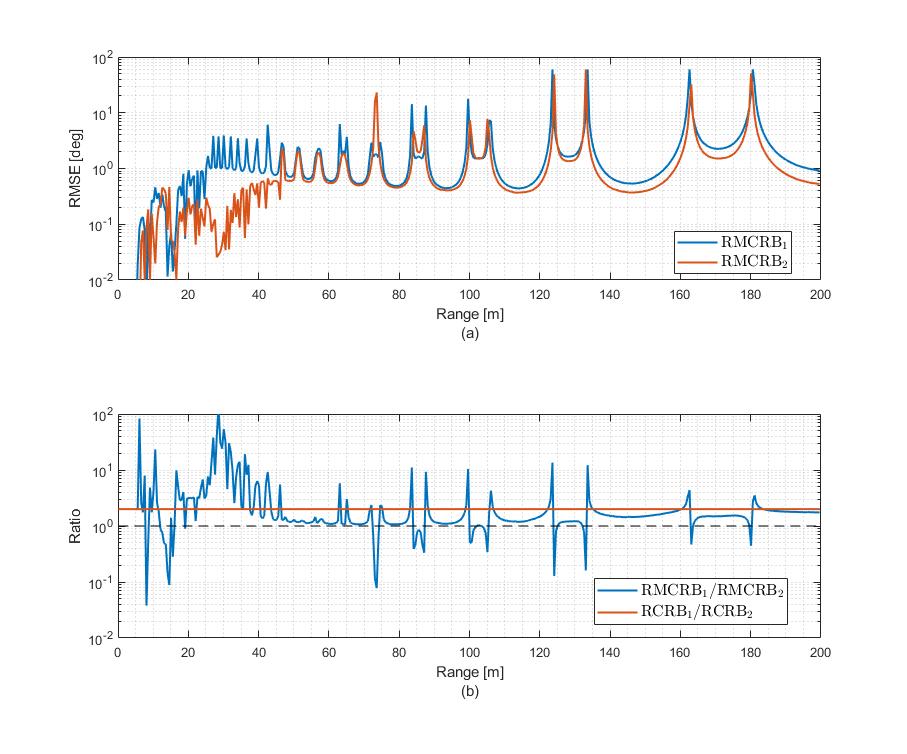}
    \caption{RMCRB and RCRB in a typical scenario with a target located at angle $\theta=0^\circ$, detected by automotive radar with two different array configurations. (a) RMCRBs for two MIMO array configurations: Blue curve represents MIMO radar array configuration with $M_t=3, M_r=8$ array elements, and the orange curve represents MIMO radar array configuration with $M_t = 3, M_r=16$ array elements. (b) Ratios between the RMCRBs and the RCRBs for the two considered MIMO arrays.}
    \label{fig:scenario_2}
\end{figure}

Fig.~\ref{fig:scenario_2} compares the RMCRB with the RCRB on target elevation DOA estimation RMSE in the automotive scenario described in Fig.~\ref{fig:scenario} as a function of the target range, $r_d$, for two considered MIMO radar configurations: a) vertical arrays of $3$ Tx and $8$ Rx  elements, and b) vertical arrays $3$ Tx and $16$ Rx elements. The RCRB and RMCRB for the array configurations (a) and (b) will be denoted as $\mathrm{RCRB_1}, \mathrm{RMCRB_1}$ and $\mathrm{RCRB_2}, \mathrm{RMCRB_2}$, respectively. 
First, notice that the RMCRB predicts high RMSE at a set of discrete ranges that correspond to ``blind'' ranges of destructive multipath when the phase difference between paths is $\frac{2k}{3}\pi$, where $k$ is an arbitrary integer. This result supports the observation in Fig.~\ref{fig:scenario_1}. 
Similarly to the result in Fig.~\ref{fig:MCRB6}, the reflection coefficient amplitude is $\left|\Gamma_r\right|\approx 1$, and the RMCRB exhibits the periodic pattern of the phase difference at long ranges. 
Notice that the phase difference change rate decreases with increasing the target range. This result is associated with the decreasing change rate of $\Delta r$, which ``stretches'' the periodic pattern across the ranges. 

It can be seen in Fig.~\ref{fig:scenario_2} that at short ranges of up to $50$ m, there is a significant difference between the two considered MIMO array configurations. This observation can be explained by the wider aperture of the array configuration with $3$ Tx and $16$ Rx array elements. This array has a narrower beamwidth, and therefore, the indirect path appears in the beampattern sidelobe ($\Delta\theta$ is larger than the main beamwidth), resulting in lower DOA estimation errors.
At long ranges, the difference between the direct and indirect path DOAs, $\Delta\phi$, decreases and the RMCRBs evaluated for the two considered MIMO array configurations coincide. Surprisingly, the ratio between the RMCRBs is lower than that between the RCRBs (at ranges beyond $40$ m). This observation intriguingly suggests that  larger arrays provide narrower beams, and thus, in short ranges, the indirect path components fall in a different beam, creating a ghost target. But for longer ranges, where  $\Delta \theta \approx 0$, the angular resolution may not be sufficient to resolve the direct and indirect components. {This observation shows that an extremely narrow beam is required for long-range radars to overcome the multipath effects. However, this requirement can not be met in practical automotive radars, limited by their size. Therefore, increasing the radar array aperture within practical dimension limitations could not significantly improve the DOA estimation performance in the considered multipath scenario.}
 
\section{Discussion}

{This work investigates the DOA estimation performance degradation due to the presence of  radar echo model misspecification, where the assumed model by the estimator does not consider the presence of multipath. Deriving estimators which consider the accurate multipath propagation model is impractical for the following reasons: 1) the multipath geometry is typically unknown in practical applications, 2) even if the model is perfectly known, the resulting DOA estimation algorithm would be significantly more computationally complex compared to the conventional approaches, since it is required to estimate the multipath parameters jointly, and 3) in some scenarios, more precise models which include additional parameters, may provide worse performance compared to misspecified models (see for example~\cite{9746339,lubeigt2023untangling}).}

{It is important to note that similar to all non-Bayesian bounds the proposed MCRB-based tool is intended for system design and offline analysis of DOA estimation performance and can not be used as an online, operational tool. This is due to the fact that the bound depends on the true model and the actual parameters, and evaluating the radar performance in a specific multipath scenario requires the consideration of particular multipath parameters. The analytical relation between direct and reflected path directions,  $\theta$ and $\psi$ in \eqref{psi_} pertains to a typical automotive scenario, as shown in Fig. \ref{fig:scenario}. However, the actual scenario may differ, and therefore, the estimator cannot rely on this knowledge. The approach presented in this paper provides a useful tool for performance analysis in  multipath-induced modeling misspecification, which can be easily evaluated in other automotive scenarios.
}

{In conventional CRB on DOA estimation with deterministic complex amplitude in the presence of additive Gaussian noise, there is no coupling between the noise covariance parameters and the target parameters. This implies that a lack of knowledge of the covariance matrix, $\mathbf{R}$, does not influence the DOA estimation performance. In MCRB, this property is not satisfied in general~\cite{7421439}. However, for diagonal covariance matrix, $\mathbf{R}=\sigma_w^2\mathbf{I}_{M_r}$, the lack of knowledge of $\sigma_w^2$ does not affect the MCRB. This result can be explained by the observation that MML is independent of $\sigma_w^2$. Therefore, knowledge of $\sigma_w^2$ does not affect the MML. Since the MML asymptotically attains the MCRB, and because the same estimator is obtained when $\sigma_w^2$ is either known or unknown, it can be deduced that the MCRB on DOA estimation is identical for both known and unknown noise covariance. In practice, considering stationary noise, $\mathbf{R}$, can be estimated using the prior measurements or based on free-target range-Doppler cells.}


\section{Conclusion}
Multipath is a major phenomenon that limits the automotive radar DOA estimation performance, mainly due to model misspecification introduced by the presence of indirect paths in addition to the direct path. In this work, we derived the MCRB on the radar DOA estimation performance in the presence of a multipath. The ability of the MCRB to accurately predict the achievable estimation performance in a typical automotive scenario was demonstrated, and it was shown that after the threshold region, it coincides with the MML estimator RMSE. The MCRB was used to investigate the multipath effects on the radar DOA estimation performance in a practical automotive scenario. The expected DOA estimation performance was studied as a function of scenario geometry and  sensor array configuration. The derived MCRB provides a useful tool for system design, such as optimal array geometry design in the presence of a multipath. 

This work can be extended to analyze the multipath effect for two-dimensional arrays and for more automotive scenarios, such as multipath induced by guard rails, bridges, and other dominant reflectors, which include both azimuth and elevation angles. New system design methods, such as array geometry optimization, can be developed based on the proposed MCRB-based tool. 

\appendices

\section{FIM Elements for MIMO radar}
In this appendix, the elements of the FIM for a multipath-free MIMO radar scenario, derived in~\cite{tabrikian2008performance}, are presented. The diagonal elements of the FIM 
are given by:
\begin{align*}
&\mathbf{J}_{\bm{\alpha}_d \bm{\alpha}_d}=\mathrm{diag}\left(\frac{2K}{\sw }\mathrm{Re}\left\{[1\; j]^H\tr(\Ad \mathbf{R}_{pp}(\tau_d)\AdH)[1 \; j]\right\}\right)\;,\\
&\mathrm{J}_{\theta \theta}=\frac{2K|\alpha_d|^2}{\sw }\mathrm{Re}\left\{\tr(\dAd \mathbf{R}_{pp}(\tau_d)\dAdH)\right\}\;, \\
&\mathrm{J}_{\tau_d \tau_d}=\frac{2K|\alpha_d|^2}{\sw }F_{\tau_d}\;,\\
&\mathrm{J}_{\od \od}=\frac{2K|\alpha_d|^2}{\sw }F_{\od}\;,\\
\end{align*}
where $\bm{\alpha}_d \triangleq [\alpha_R \;\; \alpha_I]^T$, 
\begin{align*}
&F_{\tau_d}=\mathrm{Re}\left\{\tr(\Ad \mathbf{R}_{t^2pp}(\tau_d)\AdH)\right\}\;,\\
&F_{\od}=\mathrm{Re}\left\{\tr(\dAd \mathbf{R}_{pp}(\tau_d)\dAdH)\right\}\;,
\end{align*}
and 
\begin{align*}
    &\mathbf{R}_{pp}(\tau) = \int_{-T/2}^{T/2}\p(t-\tau)\p^H(t-\tau)dt\;,\\
    &\mathbf{R}_{\dot{p}\dot{p}}(\tau) = \int_{-T/2}^{T/2}\dot{\p}(t-\tau)\dot{\p}^H(t-\tau)dt\;,\\
    &\mathbf{R}_{t^2pp}(\tau) = \int_{-T/2}^{T/2}t^2\p(t-\tau)\p^H(t-\tau)dt\;, 
\end{align*}

We consider MIMO radar with orthogonal signals, that is 
\begin{equation}\label{i1}
\int_{-T/2}^{T/2}\p(t)\p^H(t)dt = \Ep\iden_{M_t}\;.
\end{equation}
Assuming that the transmit and receive steering vectors are normalized, $\mathbf{a}_r^H\mathbf{a}_r = \mathbf{a}_t^H\mathbf{a}_t = 1$, we obtain 
\begin{align}\label{i2}
    \tr(\Ad\AdH) = \tr(\mathbf{a}_r\mathbf{a}_t^T\mathbf{a}_t^*\mathbf{a}_r^H) = 1\;.
\end{align} 
By using (\ref{i1}) and (\ref{i2}), $\mathbf{J}_{\bm{\alpha}_d \bm{\alpha}_d}$ and $\mathrm{J}_{\theta \theta}$ can be rewritten as
\begin{align}
&\mathbf{J}_{\bm{\alpha}_d \bm{\alpha}_d}=\frac{2K\Ep}{\sw }\iden_{2}\label{Jxi1}\;,\\
&\mathrm{J}_{\theta \theta}=\frac{2K\abs{\ad}^2\Ep}{\sw }\tr(\dAd \dAdH )\;.\label{Jxi2}
\end{align}
The range and Doppler elements in the FIM depend on the transmitted pulse and can be denoted as: 
\begin{align}
    &\mathrm{J}_{\tau_d \tau_d} =
    \frac{2K|\alpha_d|^2}{\sw }F_{\tau_d}\;,\\
    &\mathrm{J}_{\od \od} = \frac{2K|\alpha_d|^2}{\sw }F_{\od}\;.\label{Jxi4}   
\end{align}
{The off-diagonal elements of the FIM are given by ~\cite{tabrikian2008performance}
\begin{align}
&\mathbf{J}_{\tau_d\bm{\alpha}_d}=-\frac{2K}{\sw }\mathrm{Re}\left\{\alpha_d^*\tr(\Ad \mathbf{R}_{p\dot{p}}(\tau_d)\AdH)[1 \; j]\right\}\label{TauAlpha}\\
&\mathbf{J}_{\bm{\alpha}_d\omega_{D_d}}=-\frac{2K}{\sw }\mathrm{Re}\left\{[1 \; j]^H\alpha_d\tr(\Ad \mathbf{R}_{tpp}(\tau_d)\AdH)\right\}\label{alphaOmega}\\
&\mathbf{J}_{\theta\bm{\alpha}_d}=-\frac{2K}{\sw }\mathrm{Re}\left\{\alpha_d^*\tr(\Ad \mathbf{R}_{pp}(\tau_d)\dAdH[1 \; j])\right\}\label{thetaAlpha}\\
&\mathbf{J}_{\tau_d\omega_{D_d}}=-\frac{2|\alpha|^2K}{\sw }\mathrm{Re}\left\{j\tr(\Ad \mathbf{R}_{tp\dot{p}}(\tau_d)\AdH)\right\}\label{TauDop}\\
&\mathbf{J}_{\theta\tau_d}=-\frac{2|\alpha|^2K}{\sw }\mathrm{Re}\left\{\tr(\Ad \mathbf{R}_{p\dot{p}}(\tau_d)\dAdH)\right\}\label{thetaTau}\\
&\mathbf{J}_{\theta\omega_{D_d}}=-\frac{2|\alpha|^2K}{\sw }\mathrm{Re}\left\{j\tr(\Ad \mathbf{R}_{tpp}(\tau_d)\dAdH)\right\}\label{thetaOmega}
\end{align}
where 
\begin{align}
    &\mathbf{R}_{p\dot{p}}(\tau) = \int_{-T/2}^{T/2}\p(t-\tau)\dot{\p}^H(t-\tau)dt\;,\label{R1}\\
    &\mathbf{R}_{tpp}(\tau) = \int_{-T/2}^{T/2}t\p(t-\tau)\p^H(t-\tau)dt\;,\label{R2}\\
    &\mathbf{R}_{tp\dot{p}}(\tau) = \int_{-T/2}^{T/2}t\p(t-\tau)\dot{\p}^H(t-\tau)dt\;.
\end{align}
}
{Note that by using the orthogonality assumption in \eqref{i1}, and by applying the identity in \eqref{i3}, the FIM element in \eqref{thetaAlpha} vanishes, which implies that there is no coupling between the $(\bm{\alpha},\theta)$ pair. However, the mixed terms of $\theta$ with $\omega_{D_d}$ and with $\tau_d$ are not necessarily zero unless the signals are designed to obey orthogonality in terms of \eqref{R1} and \eqref{R2}. Under this assumption, and by using the identity in \eqref{i3}, the rest of the diagonal elements which affect the estimation of $\theta$, are zero.}




\section{Derivation of the Elements of the matrix $\Cd$}
In this appendix, expressions for the elements of the matrix $\mathbf{C}_D$ from (\ref{CDdef}) are derived.
For simplicity, let: 
\begin{align}
    &\Delta \bm{\mu}_k = \bm{\mu}_k' - \bm{\mu}_k(\xxi) = \zikn\label{deltaMu2}\;,\\
    &\alpha_R\triangleq{\mathrm{Re}\{\ad\}}\;,\\
    &\alpha_I\triangleq{\mathrm{Im}\{\ad\}}\;.
\end{align}
The diagonal elements of the FIM are listed in \eqref{Jxi1}-\eqref{Jxi4}, and thus
\begin{equation}
\label{Jxi}
    \begin{split}
    \J(\xxi)=
    \frac{2K}{\sw }\mathrm{diag}
        \scriptstyle{
        \left(\left[
            \Ep \;\; \Ep \;\; |\alpha_d|^2F_{\tau_d} \;\;|\alpha_d|^2F_{\od}  \;\; E_p\abs{\ad}^2E_{\Dot{A}}\right]\right)\;.
        }
    \end{split}
\end{equation}
Substituting \eqref{deltaMu2} and \eqref{Jxi} into \eqref{CDdef}, by summing the contribution from the $K$ statistically independent observations, the elements for $\mathbf{C}_{D}$ can be expressed as:
\begin{equation}
    \begin{bmatrix}\Cd\end{bmatrix}_{ij}=[\mathbf{J(\xxi)}]_{ij}-\frac{2}{\sw }\mathrm{Re}\left\{\sumk \int_T \frac{\partial^2\zdkn^H}{\partial \xi_i\partial\xi_j}\zikn dt \right\}\;.\label{Cterm}\nonumber
\end{equation}
Using the definition of $\xxi$, the elements of~\eqref{Cterm} can be obtained as:
\begingroup
\allowdisplaybreaks
\begin{align}
        \begin{bmatrix}\Cd\end{bmatrix}_{11}&=\frac{2K\Ep}{\sw }-{\frac{2}{\sw }\mathrm{Re}\left\{\sumk \int_T \frac{\partial^2\zdkn^H}{\partial\adR^2}\zikn dt \right\}}\nonumber
        \\&= \frac{2K\Ep}{\sw }\nonumber\\        \begin{bmatrix}\Cd\end{bmatrix}_{22}&=\frac{2K\Ep}{\sw }-{\frac{2}{\sw }\mathrm{Re}\left\{\sumk \int_T \frac{\partial^2\zdkn^H}{\partial\adI^2}\zikn dt \right\}}\nonumber\\&=
        \frac{2K\Ep}{\sw }\nonumber\\
    \begin{bmatrix}\Cd\end{bmatrix}_{33}&=\;
    \mathrm{J}_{\tau_d \tau_d}-\frac{2}{\sw }\mathrm{Re}\left\{\sumk \int_T \frac{\partial^2\zdkn^H}{\partial\tau_d ^2}\zikn dt \right\}\nonumber\\
    \begin{bmatrix}\Cd\end{bmatrix}_{44}&=\;
    \mathrm{J}_{\od \od}-\frac{2}{\sw }\mathrm{Re}\left\{\sumk \int_T \frac{\partial^2\zdkn^H}{\partial\od ^2}\zikn dt \right\}\nonumber\\
    \begin{bmatrix}\Cd\end{bmatrix}_{55}&=\;
        \frac{2K\abs{\ad}^2\Ep E_{\dot{A}}}{\sw}-\nonumber\\&\frac{2}{\sw }\mathrm{Re}\left\{\sumk \int_T \frac{\partial^2\zdkn^H}{\partial\theta ^2}\zikn dt \right\}\nonumber\\ &=\frac{2K\Ep}{\sw }(\abs{\ad}^2E_{\dot{A}}-\mathrm{Re}\left\{\ad^*\ai\tr(\ddAdH\Ai)\right\})\nonumber\\        \begin{bmatrix}\Cd\end{bmatrix}_{12}&=\begin{bmatrix}\Cd\end{bmatrix}_{21}=-{\frac{2}{\sw }\mathrm{Re}\left\{\sumk \int_T \frac{\partial^2\zdkn^H}{\partial\adR\partial\adI}\zikn dt \right\}}\nonumber\\
        &=0  \nonumber\\        \begin{bmatrix}\Cd\end{bmatrix}_{13}&=\begin{bmatrix}\Cd\end{bmatrix}_{31}=\;-\frac{2}{\sw }\mathrm{Re}\left\{\sumk \int_T \frac{\partial^2\zdkn^H}{\partial\adR\partial\tau_d }\zikn dt \right\}\nonumber\\        \begin{bmatrix}\Cd\end{bmatrix}_{14}&=\begin{bmatrix}\Cd\end{bmatrix}_{41}=\;-\frac{2}{\sw }\mathrm{Re}\left\{\sumk \int_T \frac{\partial^2\zdkn^H}{\partial\adR\partial\od }\zikn dt \right\}\nonumber\\&=
        \frac{2\T}{\sw }\mathrm{Re}\left\{ j\ai\tr(\mathbf{R}_{pp}(\Delta \tau)\AdH\Ai)\sumk k \eid\right\}\nonumber\\        \begin{bmatrix}\Cd\end{bmatrix}_{15}&=\begin{bmatrix}\Cd\end{bmatrix}_{51}=\;-\frac{2}{\sw }\mathrm{Re}\left\{\sumk \int_T \frac{\partial^2\zdkn^H}{\partial\adR\partial\theta}\zikn dt \right\}\nonumber\\&=
        -\frac{2E_p}{\sw }\mathrm{Re}\left\{\ai\tr(\dAdH\Ai)\sumk \eid \right\}\nonumber\\
        \begin{bmatrix}\Cd\end{bmatrix}_{23}&= \begin{bmatrix}\Cd\end{bmatrix}_{32}=
        -\frac{2}{\sw }\mathrm{Re}\{\sumk\int_T \frac{\partial^2\zdkn^H}{\partial\adI\partial\tau_d}\zikn dt\}\nonumber\\
        \begin{bmatrix}\Cd\end{bmatrix}_{24}&= \begin{bmatrix}\Cd\end{bmatrix}_{42}=
        -\frac{2}{\sw }\mathrm{Re}\{\sumk \int_T \frac{\partial^2\zdkn^H}{\partial\adI\partial\od}\zikn dt\}\nonumber\\&=
        -\frac{2\T}{\sw }\mathrm{Re}\left\{\ai\tr(\mathbf{R}_{pp}(\Delta \tau)\AdH\Ai) \sumk k \eid \right\}\nonumber\\
        \begin{bmatrix}\Cd\end{bmatrix}_{25}&=\begin{bmatrix}\Cd\end{bmatrix}_{52}=0\;-\frac{2}{\sw }\mathrm{Re}\left\{\sumk \int_T \frac{\partial^2\zdkn^H}{\partial\adI\partial\theta}\zikn dt \right\}\nonumber\\&=
        \frac{2}{\sw }\mathrm{Re}\left\{j\ai\tr(\mathbf{R}_{pp}(\Delta \tau)\dAdH\Ai)\sumk \eid \right\}\nonumber\\
        \begin{bmatrix}\Cd\end{bmatrix}_{34}&=\begin{bmatrix}\Cd\end{bmatrix}_{43}=
        \;-\frac{2}{\sw }\mathrm{Re}\left\{\sumk \int_T \frac{\partial^2\zdkn^H}{\partial\tau_d\partial\od}\zikn dt \right\}\nonumber\\
        \begin{bmatrix}\Cd\end{bmatrix}_{35}&=\begin{bmatrix}\Cd\end{bmatrix}_{53}=
        \;-\frac{2}{\sw }\mathrm{Re}\left\{\sumk \int_T \frac{\partial^2\zdkn^H}{\partial\tau_d\partial\theta}\zikn dt \right\}\nonumber\\
        \begin{bmatrix}\Cd\end{bmatrix}_{45}&=\begin{bmatrix}\Cd\end{bmatrix}_{54}=
        \;0-\frac{2}{\sw }\mathrm{Re}\left\{\sumk \int_T \frac{\partial^2\zdkn^H}{\partial\od\partial\theta}\zikn dt \right\}\nonumber\\&=
        \frac{2\T}{\sw }\mathrm{Re}\left\{j\tr(\mathbf{R}_{pp}(\Delta \tau)\dAdH\Ai)\sumk k\rho_k\right\}\;,\nonumber
    \end{align}   
\endgroup
where
\newcommand{\zza}{\ad^*\ai\eid}    
\newcommand{\zzb}{\tr(\Phi\AdH\Ai)}    
\newcommand{\zz}{\rho_k \tr(\Phi\AdH\Ai)}
\begin{align}
    \rho_k \triangleq \ad^*\ai\eid\;\label{phi_ro2}.\nonumber
\end{align}

\section{Derivation of $\M_{\theta \theta}$}
In this appendix, the MCRB for DOA estimation $M_{\theta \theta}$ is derived under the assumptions~\eqref{ass1} and~\eqref{ass2}. Under these assumptions, we obtain the following approximations:
\begin{eqnarray}
    &\rho_k(\Delta \omega) \triangleq \ad^*\ai\eid \approx \ad^*\ai \ne \mathrm{func}(k)  \;,\label{rhoApprox}\\
    &\mathbf{R}_{pp}(\Delta \tau) \approx E_p\iden_{M_t}\ne \mathrm{func}(\tau_d)\;.\label{rppApprox}
\end{eqnarray}
In addition, the order of the integration and derivative w.r.t. to $\tau_d$ in the $\mathbf{C}_D$ elements can be exchanged. Therefore, from~\eqref{rppApprox}, all the cross $\tau_d$ elements vanish. Furthermore, due to the symmetric summation w.r.t. $k$, and using \eqref{rhoApprox}, all the related cross $\od$ elements also vanish. Using~\eqref{rhoApprox} and~\eqref{rppApprox} in the non-zeros elements in $\mathbf{C}_D$, we obtain:
\begingroup
\allowdisplaybreaks
    \begin{align}
            \begin{bmatrix}\Cd\end{bmatrix}_{11} &= \begin{bmatrix}\Cd\end{bmatrix}_{22} =\frac{2K\Ep}{\sw }\;,\nonumber\\
            {\begin{bmatrix}\Cd\end{bmatrix}_{33}}&=\mathrm{J}_{\tau_d\tau_d}\triangleq \frac{2K\Ep}{\sw } \zeta_1\;,\nonumber\\
            {\begin{bmatrix}\Cd\end{bmatrix}_{44}}&= \frac{2K\Ep}{\sw } \zeta_2\;,\nonumber\\
            {\begin{bmatrix}\Cd\end{bmatrix}_{55}}&=\frac{2 K E_p}{\sw }({\abs{\ad}^2E_{\Dot{A}}-\ad^*\ai \tr(\ddAdH\Ai)})\nonumber\\&\triangleq\frac{2K\Ep}{\sw }\zeta_3\;,\nonumber\\
        \begin{bmatrix}\Cd\end{bmatrix}_{14}&=
        \begin{bmatrix}\Cd\end{bmatrix}_{41}=-\frac{2\T K E_p}{\sw }\mathrm{Re}\left\{j\ai\tr(\AdH\Ai) \right\}\nonumber\\&\triangleq-\frac{2K\Ep}{\sw }\mathrm{Re}\{j\zeta_4\}\;,\nonumber\\        {\begin{bmatrix}\Cd\end{bmatrix}_{24}}&=\begin{bmatrix}\Cd\end{bmatrix}_{42}=-\frac{2\T K E_p}{\sw }\mathrm{Re}\left\{\ai\tr(\AdH\Ai) \right\}\nonumber\\&\triangleq-\frac{2K\Ep}{\sw }\mathrm{Re}\{\zeta_4\}\;,\nonumber\\
        \begin{bmatrix}\Cd\end{bmatrix}_{15}&=
        \begin{bmatrix}\Cd\end{bmatrix}_{51}=-\frac{2K\Ep}{\sw }\mathrm{Re}\{\ai\tr(\dAdH\Ai)\}\nonumber\\&\triangleq-\frac{2K\Ep}{\sw }\mathrm{Re}\{\zeta_5\}\;,\nonumber\\
        {\begin{bmatrix}\Cd\end{bmatrix}_{25}}&=\begin{bmatrix}\Cd\end{bmatrix}_{52}=\frac{2K\Ep}{\sw }\mathrm{Re}\{j\ai\tr(\dAdH\Ai)\}\nonumber\\&\triangleq\frac{2K\Ep}{\sw }\mathrm{Re}\{j\zeta_5\}\;,\nonumber
    \end{align}   
\endgroup
Finally, $\mathbf{C}_D$ can be expressed as follows:
\begin{equation}
        \Cd=\frac{2K\Ep}{\sw}
        \bm{\mathrm{Z}}\;,
\end{equation}
where 
\begin{equation}
   \bm{\mathrm{Z}} \triangleq 
   \mathrm{Re}\left\{ \begin{pmatrix*}[c]
    1   &   0   &  0  &  -j\zeta_4 & -\zeta_5\\
    0   &   1   &  0  &  -\zeta_4 & j\zeta_5\\
    0   &   0   &  \zeta_1  &  0 & 0\\
    -j\zeta_4  &  -\zeta_4   &  0   &  \zeta_2 & 0\\
    -\zeta_5  &  j\zeta_5   &  0   &  0 & \zeta_3
    \end{pmatrix*}\right\}\;.
\end{equation}
The inverse of $\mathbf{C}_D$ is given by
\begin{equation}
        \Cd^{-1}=\frac{\sw}{2K\Ep}
        \bm{\mathrm{Z}}^{-1}\;.
\end{equation}
The MCRB for DOA estimation, $M_{\theta \theta}$, can be obtained using the inverse of $\mathbf{Z}$ and after a few lines of simple algebra, one obtains:
\begin{align}
    \M_{\theta\theta}&=\frac{\sw(|\ai \tr(\dAdH\Ai )|^2+|\ad|^2 E_{\Dot{A}})}{2K\Ep(2\mathrm{Re}\{\ad^*\ai \tr(\ddAdH\Ai)\}-\abs{\ad}^2E_{\Dot{A}})^2}\nonumber \\
    &=\mathrm{CRB}(\theta)\frac{E_{\Dot{A}}(| \tr(\dAdH\Ai )|^2+\mathrm{SMR} E_{\Dot{A}})}{(\mathrm{Re}\{ \tr(\ddAdH\Ai)e^{-j\Delta \phi}\}-\sqrt{\mathrm{SMR}}E_{\Dot{A}})^2}\;.    
\end{align}
\bibliography{references}
\bibliographystyle{ieeetr}
\end{document}